# A Two-Body Synergistic Theory for Adsorption and Desorption Kinetics on Surface

Kunpeng Chen[1,5], Jun Zhao[1,2,3,4†]

[1]School of Atmospheric Sciences, Sun Yat-sen University, Guangzhou, Guangdong, 510275, China

[2]Guangdong Province Key Laboratory for Climate Change and Natural Disaster Studies, and Institute of Earth Climate and Environment System, Sun Yat-sen University, Guangzhou, Guangdong, 510275, China

[3]Southern Laboratory of Ocean Science and Engineering (Guangdong, Zhuhai), Zhuhai, Guangdong, 519082, China

[4]Guangdong Provincial Observation and Research Station for Climate Environment and Air Quality Change in the Pearl River Estuary, Guangzhou, Guangdong, 510275, China

[5]Current affinition: Department of Environmental Sciences, College of Natural and Agricultural Sciences, University of California Riverside, CA, 92521, USA

A kinetic theory is proposed to elucidate complex nature of adsorption and desorption on surface and to calculate the adsorption and desorption rates in a practically simple way. This theory provides decomposition of the energy barriers and considers synergistic effects of lateral interactions between adsorbates. A concise formulation was derived for adsorption and desorption rates on surfaces covered with multi-component adsorbates. The adsorption and desorption rates are formulated by multiplying a two-body synergistic coefficient that is explicitly dependent on surface coverage, compared respectively to those from classical theories.

## I. INTRODUCTION

Adsorption and desorption are two fundamental phenomena in physical chemistry and they are essential for in-depth understanding of substance transportation between gaseous and condensed phases. The kinetic theory of adsorption and desorption is crucial to evaluate the rate of surface mass transfer which is fundamental for development of both industry and natural sciences relevant to interfacial processes.

For adsorption, the early classical Langmuir's adsorption model[1] is well-established and widely used in many applications. However, this model is only limited to describe simple cases due to the following assumptions[2]: a constant adsorption energy barrier, a proportional relationship between the adsorption rate and the number of sites unoccupied by the adsorbate, and a total omission of lateral interactions commonly found between adsorbates[3]. Large biases are hence generated when calculating the adsorption rate for complicated systems using the classical model. Later the kinetic lattice gas model (KLGM)[4, 5] was proposed to describe the complicated nature of physicochemical processes on crystal surface. However, parameterization in the KLGM is cumbersome and computationally demanding[6, 7], hindering its further applications.

For desorption, the most prevalent theory is the Polanyi-Wigner equation[8], which assumes a constant desorption energy barrier. This equation is practically measurable and widely used in temperature-programmed desorption (TPD) experiments for first-order desorption process[2]. Later another work used first-principle approaches based on density functional theory (DFT) with the generalized gradient approximation (GGA) to evaluate the lateral interactions and predicted TPD spectra with a derived equation for nonlinear desorption[9]. Since the results largely rely on DFT-GGA, this method can only obtain accurate results for simple structures on regular crystal surfaces (e.g. CO on $RuO_2(110)$[10], $O_2$ on $Ru(0001)$[11]). For more complicated systems, it may have limited applications.

In this work, we propose an adsorption theory to illustrate the kinetics of adsorption at interface, especially for cases in which the interactions between adsorbates are not negligible. This theory provides a fundamental yet practical approach for evaluation of the adsorption rate at a complicated interface.

## II. THEORY OF TWO-BODY SYNERGISTIC EFFECTS
### A. Kinetic theory for adsorption

In this theory, the rate of adsorption ($R^{ads}$) is product of the flux of the adsorbed species ($Z_w$) and the selection factor ($s$)[12],

$$R^{ads} = Z_w s, \tag{1}$$

$$Z_w = \frac{a_s p}{\sqrt{2\pi M k_B T}}, \tag{2}$$

where $a_s$ is the surface area, p is the partial pressure of the adsorbed species, M is the molecular weight of adsorbed species, $k_B$ is the Boltzmann constant and T is the temperature in Kelvin. We adopt Langmuir's theory when formulating the selection factor s: a molecule needs to overcome an energy barrier in order to be adsorbed on the surface. In Langmuir's theory, the adsorption rate is formulated as

$$R_L^{ads} = Z_w \left(1 - \sum_{i=1}^{n} \theta_i \right) e^{-\beta E^{ads}}, \tag{3}$$

where n is the number of components on the surface. In Eq. (3), the adsorption rate is proportional to the unoccupied surface area by adsorbates while the adsorption energy barrier $E^{ads}$ is assumed to be constant. However, in our theory, we consider the adsorption energy barrier as a function of the coverages ($\theta$) of the adsorbates and propose that

$$s = s_0 e^{-\beta E^{ads}(\theta_1, \theta_2, ...)}, \tag{4}$$

where $s_0$ is the sticking coefficient, a constant that also exists in the Langmuir's model[2], $\theta_i$ is the coverage of the $i^{th}$ component, β equals to $1/(k_B T)$. $s_0$ is equivalent to the transmission coefficient *kappa* employed in the conventional transition state theory (CTST) and is always assumed to be unity[13]. For simplicity, we adopt the same assumption so that the adsorption rate can be written as Eq. (5) by combination of Eqs. (1) and (4),

$$R^{ads} = Z_w e^{-\beta E^{ads}(\theta_1, \theta_2, ...)}. \tag{5}$$

The main task of the theory is to formulate the adsorption energy barrier $E^{ads}(\theta_1, \theta_2, ..., \theta_n)$, the energy that a molecule is required to overcome when it approaches the surface. Here n is the number of species at the interface and $E^{ads}(\theta_1, \theta_2, ..., \theta_n)$ can be considered as a function of the coverage of the n species. Note that the lateral interactions between species may also contribute to $E^{ads}(\theta_1, \theta_2, ..., \theta_n)$, which is referred to as synergistic energy in the following discussion. These synergistic effects largely influence the potential energy surface of the adsorption process. Hence $E^{ads}(\theta_1, \theta_2, ..., \theta_n)$ can be decomposed as

$$E^{ads}(\theta_1, \theta_2, ..., \theta_n) = \sum_{i=1}^{n} E^{ads}(\theta_i) + \frac{1}{2}\sum_{\substack{i,j=1 \\ i \neq j}}^{n} e(\theta_i, \theta_j) + \frac{1}{6}\sum_{\substack{i,j,k=1 \\ i \neq j \neq k}}^{n} e(\theta_i, \theta_j, \theta_k) + \cdots . \quad (6)$$

The first term on the right-hand side is the sum of contribution of each species, the second term is the sum of synergistic energy from interactions between any two species, and the third term is the sum of synergistic energy from any ternary interactions ($\theta_i$, $\theta_j$, $\theta_k$), and so on. Here we assume that the binary interactions play far more important roles in the total energy barrier than any higher order interactions in terms of the synergistic effects. Hence, we ignore many-body interactions from more than two species. Thus, Eq. (6) can be simplified as

$$E^{ads}(\theta_1, \theta_2, ..., \theta_n) \approx \sum_{i=1}^{n} E^{ads}(\theta_i) + \frac{1}{2}\sum_{\substack{i,j=1 \\ i \neq j}}^{n} e(\theta_i, \theta_j) . \quad (7)$$

If the surface contains only one species, the second term in Eq. (7) vanishes and the adsorption energy barrier is simply sum of those contributed by each species.

In order to quantitatively evaluate the adsorption energy barrier contributed by species i, we decompose it into three parts,

$$E^{ads}(\theta_i) = E_0^{ads} + E_{occ}(\theta_i) + \frac{1}{2}e(\theta_i, \theta_i) . \quad (8)$$

The first term ($E_0^{ads}$) is respectively the adsorption energy barrier of the i$^{th}$ species on an empty surface. The second term ($E_{occ}(\theta_i)$) donates the effects of occupation of adsorbates on the energy barrier, since adsorbates can block the adsorption of molecules. Here we term $E_{occ}(\theta_i)$ as occupation energy barrier (OEB) and assume that it is proportional to the coverage of the occupied adsorbates,

$$E_{occ}(\theta_i) = a_{occ,i}\theta_i , \quad (9)$$

where $a_{occ,i}$ is the first-order occupation coefficient for adsorption of the i$^{th}$ species. The third term is the synergistic energy between any two molecules of the i$^{th}$ species. Similarly, we only consider binary interactions and interactions for three- and higher bodies are neglected. Hence, for multi-component adsorption, the adsorption energy barrier is the sum of the empty surface term (constant), the occupation energy term, and self-synergistic energy term (Fig. 1), and thereby

$$E^{ads}(\theta_1, \theta_2, ..., \theta_n) = E_0^{ads} + \sum_{i=1}^{n} E_{occ}(\theta_i) + \frac{1}{2}\sum_{i=1}^{n} e(\theta_i, \theta_i) + \frac{1}{2}\sum_{\substack{i,j=1 \\ i \neq j}}^{n} e(\theta_i, \theta_j)$$

$$= E_0^{ads} + \sum_{i=1}^{n} a_{occ,i}\theta_i + \frac{1}{2}\sum_{i,j=1}^{n} e(\theta_i,\theta_j). \tag{10}$$

We further derive that $a_{occ,i}$ equals to $1/\beta$ based on the Langmuir's adsorption theory (APPENDIX A), so that Eq. (10) becomes

$$E^{ads}(\theta_1,\theta_2,\ldots,\theta_n) = E_0^{ads} + \frac{1}{\beta}\sum_{i=1}^{n}\theta_i + \frac{1}{2}\sum_{i,j=1}^{n} e(\theta_i,\theta_j). \tag{11}$$

**B. Kinetic theory for desorption**

According to the theory proposed by Polanyi and Wigner[8], the desorption rate is given by

$$R_{PW}^{des} = \nu\theta^{n^{des}} e^{-\beta E^{des}}. \tag{12}$$

Here $\nu$ is the frequency factor, representing the frequency of activation. $n^{des}$ is the order of desorption, $E^{des}$ is the desorption energy barrier. In our theory, the desorption rate ($R^{des}$) is also product of the flux of desorbing species ($\nu(\theta)$) and the selection factor that is depicted by Boltzmann factor,

$$R^{des} = \nu(\theta)e^{-\beta E^{des}(\theta_1,\theta_2,\ldots,\theta_n)}. \tag{13}$$

We assume that the desorption flux is proportional to the coverage, the same as the first order Polanyi-Wigner equation,

$$\nu(\theta) = \nu\theta. \tag{14}$$

However, the high order terms as well as the nonlinear terms are involved in the exponential term, i.e., the desorption energy barrier that can be decomposed as

$$E^{des}(\theta_1,\theta_2,\ldots,\theta_n) = E_0^{des} + \frac{1}{\beta}\sum_{i=1}^{n}\theta_i + \frac{1}{2}\sum_{i,j=1}^{n} e(\theta_i,\theta_j). \tag{15}$$

.

**C. Formulation for two-body synergistic energy**

It is obvious that the total adsorption and desorption energy barriers would be analytically quantified if the synergistic energy can be formulated. Note that the lateral interactions between adsorbates decrease with increase of their distance owing to the decreasing overlap of their wavefunctions. We hence consider lateral interaction as a function of distance, and term the energy induced by lateral interaction between molecules as lateral energy, $\varepsilon_{lat}$. We also assume that the formulation of lateral energy is similar to that of typically basic interactions (e.g. orientation effects, induction effects, and dispersion effects), which are inversely proportional to sixth power of distance[14]. Thus, for two-body interactions, $\varepsilon_{lat}$ is approximately proportional to the minus six orders of magnitude of the distance ($d$) between the two molecules of adsorbates,

$$\varepsilon_{lat} = \frac{\lambda}{d^6 + \eta}. \tag{16}$$

The form of Becke-Johnson damping[15] is introduced to optimize the depiction of lateral effects when $d$ asymptotically approaches zero. $\lambda$ is a coefficient relevant to lateral energy, and $\eta$ is the damping coefficient. Now assuming that *N* molecules on the surface, and the configuration is marked by $\omega$, we can select one of them as the center molecule and derive the sum of lateral energy as

$$\varepsilon_{\text{lat}}(\omega) = \sum_{p=1}^{N(N-1)/2} \frac{\lambda_p}{d_{\omega,p}^6 + \eta_p}. \tag{17}$$

Then we define an apparent lateral energy, $<\varepsilon_{lat}>$, to depict the comprehensive lateral effects of all the possible configurations by the form of canonical partition function.

$$\langle \varepsilon_{\text{lat}} \rangle = \frac{-1}{\beta} \ln \left( \int_\Omega e^{-\beta \varepsilon_{\text{lat}}(\omega)} d\omega \right). \tag{18}$$

Note that lateral energy is not the systematic energy, so it is not normalized, that is, the integral in Eq. (18) is not equal to one. Through this formulation, the configuration with a higher value of lateral energy can have a larger contribution to the comprehensive lateral effects. Since the apparent lateral energy is a function of all of the two-body distances of all the possible configurations, here we introduce a parameter x (termed as lateral-equivalent distance) to represent the "average distance" relevant to the interactions in form of Eq. (16),

$$x = \left( \frac{\langle \lambda \rangle}{\langle \varepsilon_{\text{lat}} \rangle} - \langle \eta \rangle \right)^{\frac{1}{6}}, \tag{19}$$

where

$$\langle \lambda \rangle = \frac{2}{N(N-1)} \left( \sum_{p=1}^{N(N-1)/2} \lambda_p \right), \tag{20}$$

$$\langle \eta \rangle = \frac{2}{N(N-1)} \left( \sum_{p=1}^{N(N-1)/2} \eta_p \right). \tag{21}$$

Indeed, the variable *x* is a measure to divide the configurations of adsorbates into different sets distinguished by the lateral energy, donated as $\Omega_x$, that is, for each set of $\Omega_x$, there are many configurations ($\omega_x$) shared the same value of *x* (Fig. 2). Combining Eqs. (17)-(21), we found that lateral energy is a function of configuration at a lateral-equivalent distance, i.e., $\varepsilon_{lat,x}(\omega)$, and Eqs. (18)-(19) can be rewritten as

$$\langle \varepsilon_{\text{lat}} \rangle (x) = \frac{-1}{\beta} \ln \left( \int_{\Omega_x} e^{-\beta \varepsilon_{\text{lat},x}(\omega)} d\omega \right), \tag{22}$$

$$\langle \varepsilon_{\text{lat}} \rangle (x) = \frac{\langle \lambda \rangle}{x^6 + \langle \eta \rangle}. \tag{23}$$

All the configurations with the same lateral-equivalent distance (*x*) can be termed as the lateral-equivalent configuration (LEC) hereafter. However, for a set of lateral-equivalent configurations, different configuration can induce different synergistic energy, i.e., synergistic-energy eigenvalue, and each configuration has its own probability of emergence. Thus, it is reasonable to calculate the expectation of synergistic energy ($<e>$) which is given by the sum of products of eigenvalue ($e^{LEC}$) and probability ($p^{LEC}$),

$$\langle e \rangle = \int_0^{+\infty} e^{LEC}(x) p^{LEC}(x) dx ,\tag{24}$$

where both $e^{LEC}$ and $p^{LEC}$ are a function of *x*. Usually the surface is quite huge compared to a single molecule, thus we assume that the upper bound of integral in Eq. (24) is infinite. As mentioned before, the synergistic effect is directly contributed by lateral effects, so we assume that the synergistic energy shares a form similar to lateral energy. For two-body synergistic interactions, any pair of molecules on the surface can contribute to synergistic energy so that a total of *N(N-1)/2* pairs exist on the *N*-molecule surface. The total synergistic energy is then formulated as

$$e^{LEC}(x) = \frac{N(N-1)}{2} \frac{Q}{x^6 + \eta_1} ,\tag{25}$$

where *Q* is the average coefficient of the synergistic energy attributed to each pair of adsorbates, and $\eta_1$ is the damping coefficient of synergistic energy. As for $p^{LEC}$, it can be formulated as

$$p^{LEC}(x) = \int_{\Omega_x} p_x(\omega) d\omega ,\tag{26}$$

where

$$p_x(\omega) = \frac{e^{-\beta \varepsilon_x(\omega)}}{\int_X \int_{\Omega_x} e^{-\beta \varepsilon_x(\omega)} d\omega dx} ,\tag{27}$$

where X, defined in Eq. (19), is the range of x, i.e., (0,+∞), and $\varepsilon_x(\omega)$ is the systematic energy. Due to the normalization of probability,

$$\int_X \int_{\Omega_x} e^{-\beta \varepsilon_x(\omega)} d\omega dx = 1 .\tag{28}$$

Eq. (27) can be simplified as

$$p_x(\omega) = e^{-\beta \varepsilon_x(\omega)} .\tag{29}$$

Note that here the system contains only the *N* molecules. For the molecules that have not been adsorbed, they are assumed to suspend above the surface and have no interactions with the adsorbates. The energy of the surface itself is set to be the zero point. Hence, for the molecules on the surface, the systematic energy can be divided into two terms: total occupation energy ($\varepsilon_{occ}$) and the lateral energy ($\varepsilon_{lat,x}(\omega)$).

$$\varepsilon_x(\omega) = \left( \sum_{p=1}^N \varepsilon_{occ,p} \right) + \varepsilon_{lat,x}(\omega) .\tag{30}$$

We assume that the lateral energy is mainly contributed by a part of electrons of the adsorbate, and these electrons mainly distribute within the wavefunctions that are far from the center of the adsorbate. Thus, it is reasonable to divide the adsorbate into two layers: the frozen core and the outer layer. The former induces occupation energy while the latter induces lateral energy. Hence, we use the product of total number of adsorbates and the average energy of the frozen cores to represent the total frozen-core energy.

$$\langle \varepsilon_{occ} \rangle = \frac{1}{N} \sum_{p=1}^N \varepsilon_{occ,p} .\tag{31}$$

We denote $\langle\varepsilon_{occ}\rangle$ as the average energy of the occupation energy. Hence, Eq. (30) becomes

$$\varepsilon_x(\omega) = N\langle\varepsilon_{occ}\rangle + \varepsilon_{lat,x}(\omega) \ . \tag{32}$$

Combining Eqs. (26) and (29), we derive

$$p^{LEC}(x) = \int_{\Omega_x} e^{-\beta[N\langle\varepsilon_{occ}\rangle + \varepsilon_{lat,x}(\omega)]} d\omega,$$

$$= e^{-\beta N\langle\varepsilon_{occ}\rangle} \int_{\Omega_x} e^{-\beta\varepsilon_{lat,x}(\omega)} d\omega. \tag{33}$$

Similarly, combining Eqs. (22) and (23), we can further derive

$$p^{LEC}(x) = e^{-\beta N\langle\varepsilon_{occ}\rangle} e^{-\beta\langle\varepsilon_{lat}\rangle(x)},$$

$$= e^{-\beta[N\langle\varepsilon_{occ}\rangle + \langle\varepsilon_{lat}\rangle(x)]},$$

$$= e^{-\beta\left[N\langle\varepsilon_{occ}\rangle + \left(\frac{\langle\lambda\rangle}{x^6 + \langle\eta\rangle}\right)\right]}. \tag{34}$$

If $\eta_2$ is used to donate $\langle\eta\rangle$, the expectation of synergistic energy can be formulated as

$$\langle e\rangle = \int_0^{+\infty} \frac{Q[N(N-1)/2]}{x^6 + \eta_1} e^{-\beta\left[N\langle\varepsilon_{occ}\rangle + \left(\frac{\langle\lambda\rangle}{x^6 + \eta_2}\right)\right]} dx \ . \tag{35}$$

Detailed derivation can be found in APPENDIX B, the final formulation can be derived as

$$\langle e\rangle = \frac{\pi}{6} Q e^{-\beta\langle\lambda\rangle/(\eta_2-\eta_1)} a_s\Theta(a_s\Theta - 1) e^{-\beta a_s\Theta\langle\varepsilon_{occ}\rangle},$$

$$= A\Theta(a_s\Theta - 1) e^{-B\Theta} \quad (A, B>0). \tag{36}$$

where

$$A = \frac{\pi}{6} Q e^{-\beta\langle\lambda\rangle/(\eta_2-\eta_1)} a_s, \quad B = \beta\langle\varepsilon_{occ}\rangle a_s. \tag{37}$$

$\Theta$ is the total coverage of the two-body synergistic components, that is, for the i$^{th}$ and j$^{th}$ component on the surface,

$$\Theta = \Theta_{ij} = \theta_i + \theta_j. \tag{38}$$

Note that A and B are coefficients relevant to a particular two-body pair and here we donate them as $A_{ij}$ and $B_{ij}$, respectively. Finally, an explicitly analytical form for the total adsorption energy barrier is derived

$$E^{ads}(\theta_1, \theta_2, \ldots, \theta_n) = E_0^{ads} + \frac{1}{\beta}\sum_{i=1}^{n}\theta_i + \frac{1}{2}\sum_{i,j=1}^{n}\langle e\rangle_{ij}$$

$$= E_0^{ads} + \frac{1}{\beta}\sum_{i=1}^{n}\theta_i + \frac{1}{2}\sum_{i,j=1}^{n}\left[A_{ij}\Theta_{ij}(a_s\Theta_{ij}-1)e^{-B_{ij}\Theta_{ij}}\right]. \tag{39}$$

## III. KINETICS OF ADSORPTION AND DESORPTION

In summary, we can express the rate of adsorption and desorption in an analytical form as

$$R^{ads} = Z_w\, e^{-\beta\left\{E_0^{ads} + (1/\beta)\sum_{i=1}^{n}\theta_i + \frac{1}{2}\sum_{i,j=1}^{n}\left[A_{ij}\Theta_{ij}\left(a_s\Theta_{ij}-1\right)e^{-B_{ij}\Theta_{ij}}\right]\right\}}, \quad (40)$$

$$R^{des} = \nu\theta\, e^{-\beta\left\{E_0^{des} + (1/\beta)\sum_{i=1}^{n}\theta_i + \frac{1}{2}\sum_{i,j=1}^{n}\left[A_{ij}\Theta_{ij}\left(a_s\Theta_{ij}-1\right)e^{-B_{ij}\Theta_{ij}}\right]\right\}}. \quad (41)$$

According to the Taylor expansion, we retain the first-order term as an approximation,

$$e^{-\sum_{i=1}^{n}\theta_i} \approx 1 - \sum_{i=1}^{n}\theta_i. \quad (42)$$

The adsorption rate is derived as

$$R^{ads} = Z_w\left(1 - \sum_{i=1}^{n}\theta_i\right)e^{-\beta E_0^{ads}}\, e^{-\frac{\beta}{2}\sum_{i,j=1}^{n}\left[A_{ij}\Theta_{ij}\left(a_s\Theta_{ij}-1\right)e^{-B_{ij}\Theta_{ij}}\right]}. \quad (43)$$

Here we define $\xi$ as the total two-body synergistic coefficient,

$$\xi = e^{-\frac{\beta}{2}\sum_{i,j=1}^{n}\left[A_{ij}\Theta_{ij}\left(a_s\Theta_{ij}-1\right)e^{-B_{ij}\Theta_{ij}}\right]}, \quad (44)$$

Thus, combining Eq. (3) of Langmuir's theory, the adsorption rate can be written as

$$R^{ads} = \xi R_L^{ads}. \quad (45)$$

As for desorption rate,

$$R^{des} = \nu\theta\left(1 - \sum_{i=1}^{n}\theta_i\right)e^{-\beta E_0^{des}}\, e^{-\frac{\beta}{2}\sum_{i,j=1}^{n}\left[A_{ij}\Theta_{ij}\left(a_s\Theta_{ij}-1\right)e^{-B_{ij}\Theta_{ij}}\right]}$$

$$= \nu\left(\theta - \theta\sum_{i=1}^{n}\theta_i\right)e^{-\beta E_0^{des}}\, e^{-\frac{\beta}{2}\sum_{i,j=1}^{n}\left[A_{ij}\Theta_{ij}\left(a_s\Theta_{ij}-1\right)e^{-B_{ij}\Theta_{ij}}\right]}. \quad (46)$$

If the coverages of studied component (i.e. $\theta$) is very small, we can only retain the first-order term and combine with Eq. (12) of Polanyi-Wigner equation, so that

$$R^{des} \approx \nu\theta\, e^{-\beta E_0^{des}}\, e^{-\frac{\beta}{2}\sum_{i,j=1}^{n}\left[A_{ij}\Theta_{ij}\left(a_s\Theta_{ij}-1\right)e^{-B_{ij}\Theta_{ij}}\right]},$$

$$= \xi R_{PW}^{des}. \quad (47)$$

This suggested that Polanyi-Wigner equation may have limited applications on high-coverage surface. Finally, Eq. (44) can be written as

$$\xi = e^{-\frac{\beta}{2} \sum_{i,j=1}^{n} \left[ A_{ij}\Theta_{ij}(a_s\Theta_{ij}-1)e^{-B_{ij}\Theta_{ij}} \right]}$$

$$= \prod_{i,j=1}^{n} e^{-\frac{\beta}{2}\left[ A_{ij}\Theta_{ij}(a_s\Theta_{ij}-1)e^{-B_{ij}\Theta_{ij}} \right]}$$

$$= \prod_{i,j=1}^{n} \xi_{ij}. \tag{48}$$

The above equation shows that the total two-body synergistic coefficient is product of each pair two-body synergistic coefficient ($\xi_{ij}$).

## IV. Conclusion

In this work, we propose a kinetic theory for adsorption and desorption that includes two-body synergistic effects on the energy barriers. The theory provides decomposition of activation energy barriers and derives analytical adsorption and desorption rates on surface where lateral effects of adsorbates cannot be neglected. The adsorption and desorption rates based on this theory are approximately equivalent to those from Langmuir's theory and the first-order Polanyi-Wigner's theory respectively multiplied by a two-body synergistic coefficient $\xi$. In principle, $\xi$ can be determined from its component ($\xi_{ij}$) which is a function of surface coverage. However, since $\kappa_{ij}$ is a nonlinear function, this parameter should be determined by nonlinear fitting of the data obtained directly from experimental measurements. Moreover, application of this theory is not restricted by the surface types and phase states (i.e., solid and droplet). It is concluded that the theory proposed in this work can provide simple but robust formulation of absorption and adsorption rates to facilitate its practical application in future surface kinetic studies.

[†] Email: zhaojun23@mail.sysu.edu.cn

**ACKNOWLEDGEMENT**

J.Z. acknowledges support from National Natural Science Foundation of China (NSFC) (91644225, 21577177, 41775117), Science and Technology Innovation Committee of Guangzhou (201803030010), the "111 plan" Project of China (Grant B17049), and Scientific and Technological Innovation Team Project of Guangzhou Joint Research Center of Atmospheric Sciences, China Meteorological Administration (Grant No.201704). K.P.C. acknowledges a special fund from the Cultivation of Guangdong Scientific and Technological Innovation Program for College Students (pdjh2017b0013).

## APPENDIX A: DERIVATION OF THE FIRST-ORDER OCCUPATION COEFFICIENT

Since Langmuir's theory can successfully describe the adsorption process for one adsorbate with low coverage, here the species occupation plays a predominant role whereas self-synergistic effects contribute insignificantly to the process. Hence the synergistic term vanishes and

adsorption rate becomes

$$R_i^{ads} = Z_w \, e^{-\beta\left(E_{i,0}^{ads}+E_{i,occ}^{ads}(\theta_i)+e_{ii}^{ads}(\theta_i)\right)}$$

$$= Z_w \, e^{-\beta\left(E_{i,0}^{ads}+E_{i,occ}^{ads}(\theta_i)\right)}$$

$$= Z_w \, e^{-\beta E_{i,0}^{ads}} \, e^{-\beta E_{i,occ}^{ads}(\theta_i)}$$

$$\approx Z_w \left[1 - \beta a_{i,occ}^{ads} \theta_i \right] e^{-\beta E_{i,occ}^{ads}(\theta_i)}. \tag{A1}$$

This is equivalent to Langmuir's adsorption model (Eq. (3)) only if

$$\beta a_{i,occ}^{ads} = 1 \quad \text{i.e.} \quad a_{i,occ}^{ads} = \frac{1}{\beta}. \tag{A2}$$

**APPENDIX B: DERIVATION OF TWO-BODY SYNERGISTIC ENERGY**

For the integral of synergistic energy,

$$\langle e \rangle = \int_0^{+\infty} \frac{Q[N(N-1)/2]}{x^6+\eta_1} e^{-\beta\left[N\langle\varepsilon_{occ}\rangle+\left(\frac{\langle\lambda\rangle}{x^6+\eta_2}\right)\right]} dx$$

$$= \frac{Q}{2} N(N-1) e^{-\beta N\langle\varepsilon_{occ}\rangle} \int_0^{+\infty} \frac{1}{x^6+\eta_1} e^{-\left(\frac{\beta\langle\lambda\rangle}{x^6+\eta_2}\right)} dx. \tag{C1}$$

We donate the integral as *I*,

$$I = \int_0^{+\infty} \frac{1}{x^6+\eta_1} e^{-\left(\frac{\beta\langle\lambda\rangle}{x^6+\eta_2}\right)} dx. \tag{C2}$$

and extend the integral to the complex plane and change the self-variable as

$$z = x + iy, (x \in \mathfrak{R}, Y \in \mathfrak{R}). \tag{C3}$$

Select the half-cycle above the *x* dimension and the *x* dimension itself as a contour (*C*) (Fig. B1) for the complex integral given by

$$J = \oint_C \varphi(z) dz$$

$$= \lim_{R \to +\infty}\left[\int_{C_R} \varphi(z) dz + \int_{-R}^{+R} \varphi(z) dz\right], \tag{C4}$$

where

$$\varphi(z) = \frac{1}{z^6+\eta_1} e^{-\left(\frac{\beta\langle\lambda\rangle}{z^6+\eta_2}\right)}. \tag{C5}$$

$C_R$ is the arc whose radium is *R*. The singularities of *φ(z)* can be obtained by

$$z^6+\eta_1 = 0 \quad \text{and} \quad z^6+\eta_2 = 0, \tag{C6}$$

and roots are listed below

$$z_1 = \sqrt[6]{\eta_1}e^{i\frac{\pi}{6}}, \quad z_2 = \sqrt[6]{\eta_1}e^{i\frac{3\pi}{6}}, \quad z_3 = \sqrt[6]{\eta_1}e^{i\frac{5\pi}{6}},$$

$$z_4 = \sqrt[6]{\eta_1}e^{i\frac{7\pi}{6}}, \quad z_5 = \sqrt[6]{\eta_1}e^{i\frac{9\pi}{6}}, \quad z_6 = \sqrt[6]{\eta_1}e^{i\frac{11\pi}{6}},$$

$$z_7 = \sqrt[6]{\eta_2}e^{i\frac{\pi}{6}}, \quad z_8 = \sqrt[6]{\eta_2}e^{i\frac{3\pi}{6}}, \quad z_9 = \sqrt[6]{\eta_2}e^{i\frac{5\pi}{6}},$$

$$z_{10} = \sqrt[6]{\eta_2}e^{i\frac{7\pi}{6}}, \quad z_{11} = \sqrt[6]{\eta_2}e^{i\frac{9\pi}{6}}, \quad z_{12} = \sqrt[6]{\eta_2}e^{i\frac{11\pi}{6}}. \tag{C7}$$

Here only six singularities ($z_1$, $z_2$, $z_3$, $z_7$, $z_8$, $z_9$) are circled by the contour. Thereby, according to the residue theorem, J can be calculated by

$$J = 2\pi i \left[\text{res}(\varphi(z_1)) + \text{res}(\varphi(z_2)) + \text{res}(\varphi(z_3)) + \text{res}(\varphi(z_7)) + \text{res}(\varphi(z_8)) + \text{res}(\varphi(z_9))\right]. \tag{C8}$$

The first residue is calculated by

$$\text{res}(\varphi(z_1)) = \lim_{z \to z_1}\left[(z-z_1)\cdot\varphi(z)\right]$$

$$= \lim_{z \to z_1}\left[(z-z_1)\frac{e^{-\left(\frac{\beta\langle\lambda\rangle}{z^6+\eta_2}\right)}}{(z-z_1)(z-z_2)(z-z_3)(z-z_4)(z-z_5)(z-z_6)}\right]$$

$$= \frac{e^{-\beta\langle\lambda\rangle/(z_1^6+\eta_2)}}{(z_1-z_2)(z_1-z_3)(z_1-z_4)(z_1-z_5)(z_1-z_6)}$$

$$= \frac{e^{-\beta\langle\lambda\rangle/(\eta_2-\eta_1)}}{(z_1-z_2)(z_1-z_3)(z_1-z_4)(z_1-z_5)(z_1-z_6)}. \tag{C9}$$

Similarly, we can also derive that

$$\text{res}(\varphi(z_2)) = \frac{e^{-\beta\langle\lambda\rangle/(\eta_2-\eta_1)}}{(z_2-z_1)(z_2-z_3)(z_2-z_4)(z_2-z_5)(z_2-z_6)}. \tag{C10}$$

$$\text{res}(\varphi(z_3)) = \frac{e^{-\beta\langle\lambda\rangle/(\eta_2-\eta_1)}}{(z_3-z_1)(z_3-z_2)(z_3-z_4)(z_3-z_5)(z_3-z_6)}. \tag{C11}$$

For simplicity, we define that

$$g(z_k) = \prod_{\substack{k=1 \\ p \neq k}}^{6}(z_k-z_p)^{-1}, \text{ (k=1,2,3)}, \tag{C12}$$

and hence

$$\text{res}(\varphi(z_k)) = e^{-\beta\langle\lambda\rangle/(\eta_2-\eta_1)}g(z_k), \text{ (k=1,2,3)}. \tag{C13}$$

As for the function $g(z_1)$,

$$g(z_1) = \frac{1}{(z_1-z_2)(z_1-z_3)(z_1-z_4)(z_1-z_5)(z_1-z_6)}$$

$$= \frac{z_1^{-5}}{(1-z_2/z_1)(1-z_3/z_1)(1-z_4/z_1)(1-z_5/z_1)(1-z_6/z_1)}$$

$$= \frac{z_1^{-5}}{\left(1-e^{i\frac{2}{6}\pi}\right)\left(1-e^{i\frac{4}{6}\pi}\right)\left(1-e^{i\frac{6}{6}\pi}\right)\left(1-e^{i\frac{8}{6}\pi}\right)\left(1-e^{i\frac{10}{6}\pi}\right)}$$

$$= \frac{z_1^{-5}}{\left(1-e^{i\frac{1}{3}\pi}\right)\left(1-e^{i\frac{2}{3}\pi}\right)(1-e^{i\pi})\left(1+e^{i\frac{1}{3}\pi}\right)\left(1+e^{i\frac{2}{3}\pi}\right)}$$

$$= \frac{z_1^{-5}}{\left(1-e^{i\frac{2}{3}\pi}\right)(1+1)\left(1-e^{i\frac{4}{3}\pi}\right)}$$

$$= \frac{z_1^{-5}}{2\left(1-e^{i\frac{2}{3}\pi}\right)\left(1-e^{-i\frac{2}{3}\pi}\right)}$$

$$= \frac{z_1^{-5}}{2\left(2-2\cos\frac{2\pi}{3}\right)}$$

$$= \frac{1}{6}z_1^{-5}. \tag{C14}$$

Similarly,

$$g(z_2) = \frac{z_2^{-5}}{(1-z_1/z_2)(1-z_3/z_2)(1-z_4/z_2)(1-z_5/z_2)(1-z_6/z_2)}$$

$$= \frac{z_2^{-5}}{\left(1-e^{-i\frac{2}{6}\pi}\right)\left(1-e^{i\frac{2}{6}\pi}\right)\left(1-e^{i\frac{4}{6}\pi}\right)\left(1-e^{i\frac{6}{6}\pi}\right)\left(1-e^{i\frac{8}{6}\pi}\right)}$$

$$= \frac{z_2^{-5}}{2\left(1-e^{-i\frac{2}{3}\pi}\right)\left(1-e^{i\frac{2}{3}\pi}\right)}$$

$$= \frac{1}{6}z_2^{-5}. \tag{C15}$$

$$g(z_3) = \frac{z_3^{-5}}{(1-z_1/z_3)(1-z_2/z_3)(1-z_4/z_3)(1-z_5/z_3)(1-z_6/z_3)}$$

$$= \frac{z_3^{-5}}{\left(1-e^{-i\frac{4}{6}\pi}\right)\left(1-e^{-i\frac{2}{6}\pi}\right)\left(1-e^{i\frac{2}{6}\pi}\right)\left(1-e^{i\frac{4}{6}\pi}\right)\left(1-e^{i\frac{6}{6}\pi}\right)}$$

$$= \frac{z_3^{-5}}{2\left(1-e^{i\frac{2}{3}\pi}\right)\left(1-e^{-i\frac{2}{3}\pi}\right)}$$

$$= \frac{1}{6} z_3^{-5}. \tag{C16}$$

Thus,

$$\text{res}(\varphi(z_1)) = \frac{1}{6} e^{-\beta\langle\lambda\rangle/(\eta_2-\eta_1)} z_1^{-5}, \tag{C17}$$

$$\text{res}(\varphi(z_2)) = \frac{1}{6} e^{-\beta\langle\lambda\rangle/(\eta_2-\eta_1)} z_2^{-5}, \tag{C18}$$

$$\text{res}(\varphi(z_3)) = \frac{1}{6} e^{-\beta\langle\lambda\rangle/(\eta_2-\eta_1)} z_3^{-5}. \tag{C19}$$

The other three singular points ($z_7$, $z_8$, $z_9$) are all removable singularities. Taking $z_7$ as an example,

$$\lim_{z \to z_7} \varphi(z) = \frac{1}{z_7^6 + \eta_1} e^{-\left(\frac{\beta\langle\lambda\rangle}{z_7^6+\eta_2}\right)} = \frac{1}{\eta_1-\eta_2} e^{-\left(\frac{\beta\langle\lambda\rangle}{\eta_2-\eta_2}\right)} = \frac{e^{-\infty}}{\eta_1-\eta_2} = 0, \tag{C20}$$

The case of $z_8$ and $z_9$ are the same. Hence, according to the definition of singularity in complex analysis,

$$\text{res}(\varphi(z_7)) = \text{res}(\varphi(z_8)) = \text{res}(\varphi(z_9)) = 0. \tag{C21}$$

Therefore, the countor integual J is solved,

$$J = 2\pi i \left[ \frac{1}{6} e^{-\beta\langle\lambda\rangle/(\eta_2-\eta_1)} \left(z_1^{-5} + z_2^{-5} + z_3^{-5}\right) + 0 + 0 + 0 \right]$$

$$= i\frac{\pi}{3} e^{-\beta\langle\lambda\rangle/(\eta_2-\eta_1)} \left(z_1^{-5} + z_2^{-5} + z_3^{-5}\right)$$

$$= i\frac{\pi}{3} e^{-\beta\langle\lambda\rangle/(\eta_2-\eta_1)} \left(e^{-i\frac{5}{6}\pi} + e^{-i\frac{15}{6}\pi} + e^{-i\frac{25}{6}\pi}\right)$$

$$= i\frac{\pi}{3} e^{-\beta\langle\lambda\rangle/(\eta_2-\eta_1)} \left(-e^{-i\frac{1}{6}\pi} + e^{-i\frac{1}{2}\pi} + e^{i\frac{1}{6}\pi}\right)$$

$$= i\frac{\pi}{3} e^{-\beta\langle\lambda\rangle/(\eta_2-\eta_1)} \left(-2i\sin\frac{\pi}{6} - i\right)$$

$$= \frac{2\pi}{3} e^{-\beta\langle\lambda\rangle/(\eta_2-\eta_1)}. \tag{C22}$$

Combining with Eq. (C4), we can derive that

$$\lim_{R \to +\infty} \left[ \int_{C_R} \varphi(z)dz + \int_{-R}^{+R} \varphi(z)dz \right] = \frac{2\pi}{3} e^{-\beta\langle\lambda\rangle/(\eta_2-\eta_1)}, \tag{C23}$$

That is,

$$\lim_{R \to +\infty} \left[ \int_{C_R} \varphi(z)dz \right] + \int_{-\infty}^{+\infty} \varphi(x)dx = \frac{2\pi}{3} e^{-\beta\langle\lambda\rangle/(\eta_2-\eta_1)}. \tag{C24}$$

Note that $\varphi(x)$ is an even function, since when $x>0$,

$$\varphi(-x) = \frac{1}{(-x)^6 + \eta_1} e^{-\left(\frac{\beta\langle\lambda\rangle}{(-x)^6+\eta_2}\right)} = \frac{1}{x^6 + \eta_1} e^{-\left(\frac{\beta\langle\lambda\rangle}{x^6+\eta_2}\right)} = \varphi(x), \tag{C25}$$

Thereby $\int_{-\infty}^{+\infty} \varphi(x)dx = 2\int_{0}^{+\infty} \varphi(x)dx$. Combining with Eq. (C24), we can derive that

$$\lim_{R \to +\infty} \left[ \int_{C_R} \varphi(z)dz \right] + 2I = \frac{2\pi}{3} e^{-\beta\langle\lambda\rangle/(\eta_2-\eta_1)}. \tag{C26}$$

To solve the limit, we firstly define another complex function $\phi(z)$,

$$\phi(z) = \varphi(z)e^{-ipz}, (p>0). \tag{C27}$$

Since

$$\lim_{z \to +\infty} \phi(z) = \lim_{z \to +\infty} \frac{1}{z^6 + \eta_1} e^{-\left(\frac{\beta\langle\lambda\rangle}{z^6+\eta_2}+ipz\right)} = 0, \tag{C28}$$

according to Jordan's Lemma,

$$\lim_{R \to +\infty} \left[ \int_{C_R} \phi(z)e^{ipz}dz \right] = 0, \text{ i.e. } \lim_{R \to +\infty} \left[ \int_{C_R} \varphi(z)dz \right] = 0. \tag{C29}$$

Therefore, we obtain the solution of $I$ and $<e>$,

$$I = \frac{\pi}{3} e^{-\beta\langle\lambda\rangle/(\eta_2-\eta_1)}, \tag{C30}$$

$$\langle e \rangle = \frac{\pi}{6} Q e^{-\beta\langle\lambda\rangle/(\eta_2-\eta_1)} N(N-1) e^{-\beta N\langle\varepsilon_{occ}\rangle}. \tag{C31}$$

Finally, we can calculate N as the product of the surface area ($a_s$) and total coverage ($\Theta$),

$$N = a_s \Theta. \tag{C32}$$

Hence,

$$\langle e \rangle = \frac{\pi}{6} Q e^{-\beta\langle\lambda\rangle/(\eta_2-\eta_1)} a_s \Theta (a_s \Theta - 1) e^{-\beta a_s \Theta \langle\varepsilon_{occ}\rangle}. \tag{C33}$$

Figure 1 Schematic representation of decomposition of activation energy barrier in our theory.

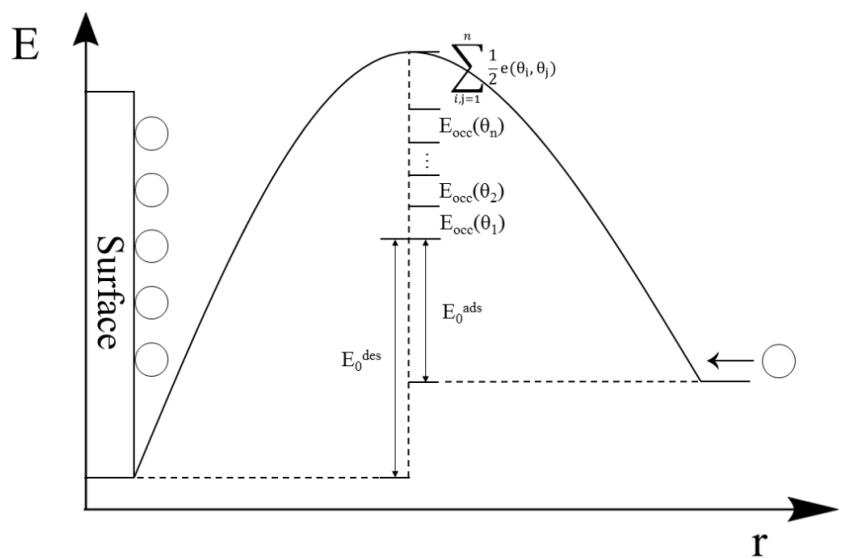

Figure 2 Schematic graph of the reflection by lateral-equivalent distance $x$.

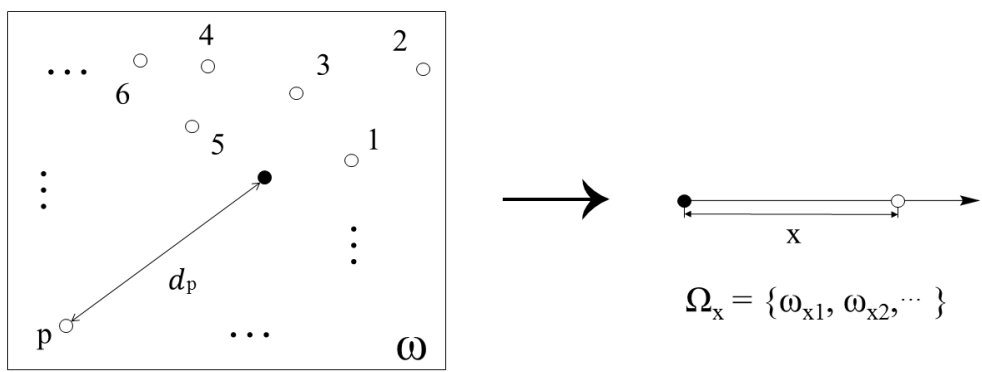

Figure B1 Contour of the complex integral.

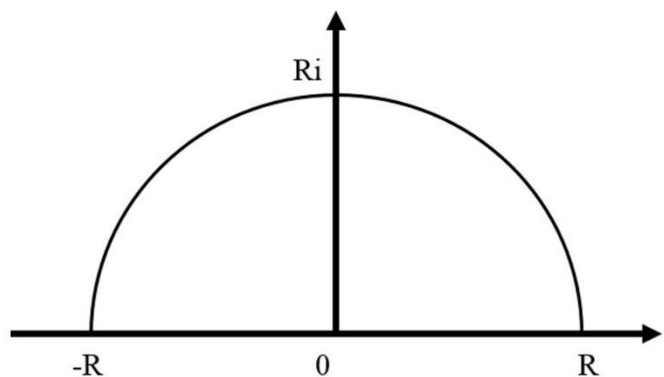